\documentclass[conference]{IEEEtran}

\ifCLASSINFOpdf
  \usepackage[pdftex]{graphicx}
  \usepackage{subfigure}
  \usepackage{float} 
\else
\fi

\begin{document}
%
\title{A Bioimpedance Measurement System for Pulse Wave Analysis}

\author{\IEEEauthorblockN{R. Kusche, T. D. Adornetto, P. Klimach and M. Ryschka}
\IEEEauthorblockA{Laboratory of Medical Electronics (LME)\\
L\"ubeck University of Applied Sciences\\
L\"ubeck, Germany\\
Email: Roman.Kusche@fh-luebeck.de, Ryschka@fh-luebeck.de}
}

\maketitle

\begin{abstract}
The morphology and velocity of the pulse wave in the arteries provide meaningful information about the cardiovascular system. Nowadays, the pulse wave is usually acquired using common blood pressure cuffs at the extremities.\\
This work describes a new measurement device which detects changes in bioimpedance due to the pulse wave. The system is able to measure up to 1000~bioimpedances per second with a resolution of 24~bits. For comparison purposes, an additional synchronized Photoplethysmography circuit is implemented. Initial measurements proved that the system is able to acquire the pulse wave at the forearm.

\end{abstract}

\vspace{2mm}\textit{Keywords--- Bioimpedance, Pulse Wave (PW), Arterial Stiffness, Photoplethysmography (PPG)}

%
\IEEEpeerreviewmaketitle

\section{Introduction}

Pulse wave analysis has proven to be an important factor in determining arterial stiffness \cite{guerin,laurent1,laurent2}. Common systems used to acquire the pulse wave are based upon pressure measurements taken at the subject's extremities \cite{Avolio}. This strategy, however, is unable to record prolonged measurements and can cause discomfort to the subject. 
\\
This paper describes a non-invasive bioimpedance measurement system capable of quantizing a pulse wave at 1000~bioimpedances per second with a resolution of up to 24~bits. 
The method is painless and has no known hazards.

\section{SYSTEM DESIGN}

\begin{figure}[H]
	\centering
	\includegraphics[width = 0.50\textwidth]{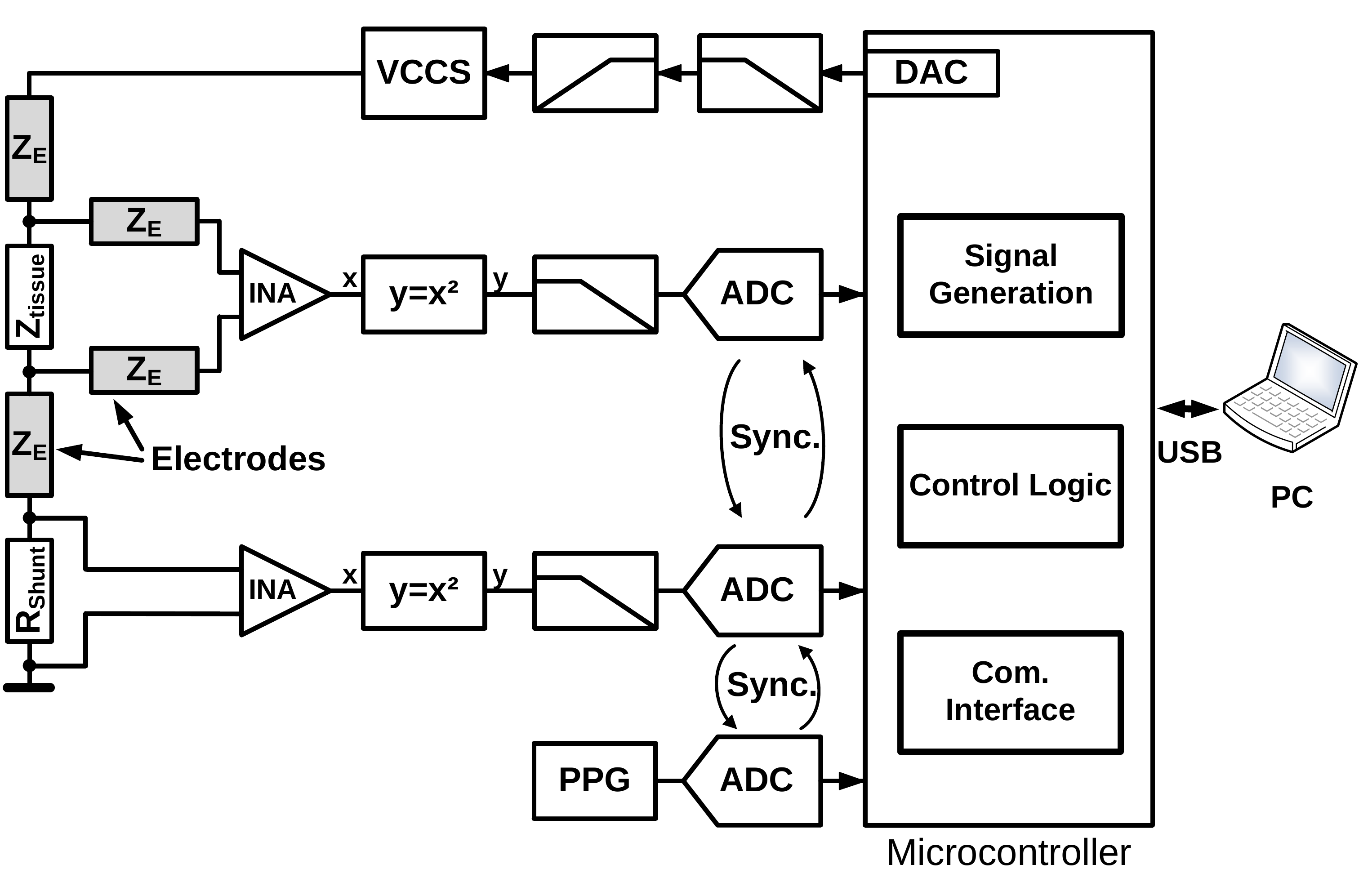}
	\caption{Principle block diagram of the developed microcontroller based bioimpedance measurement system for acquiring pulse waves.}
	\label{fig:blockDiagram}
\end{figure}

\begin{figure}[H]
	\centering
	\includegraphics[width = 0.485\textwidth]{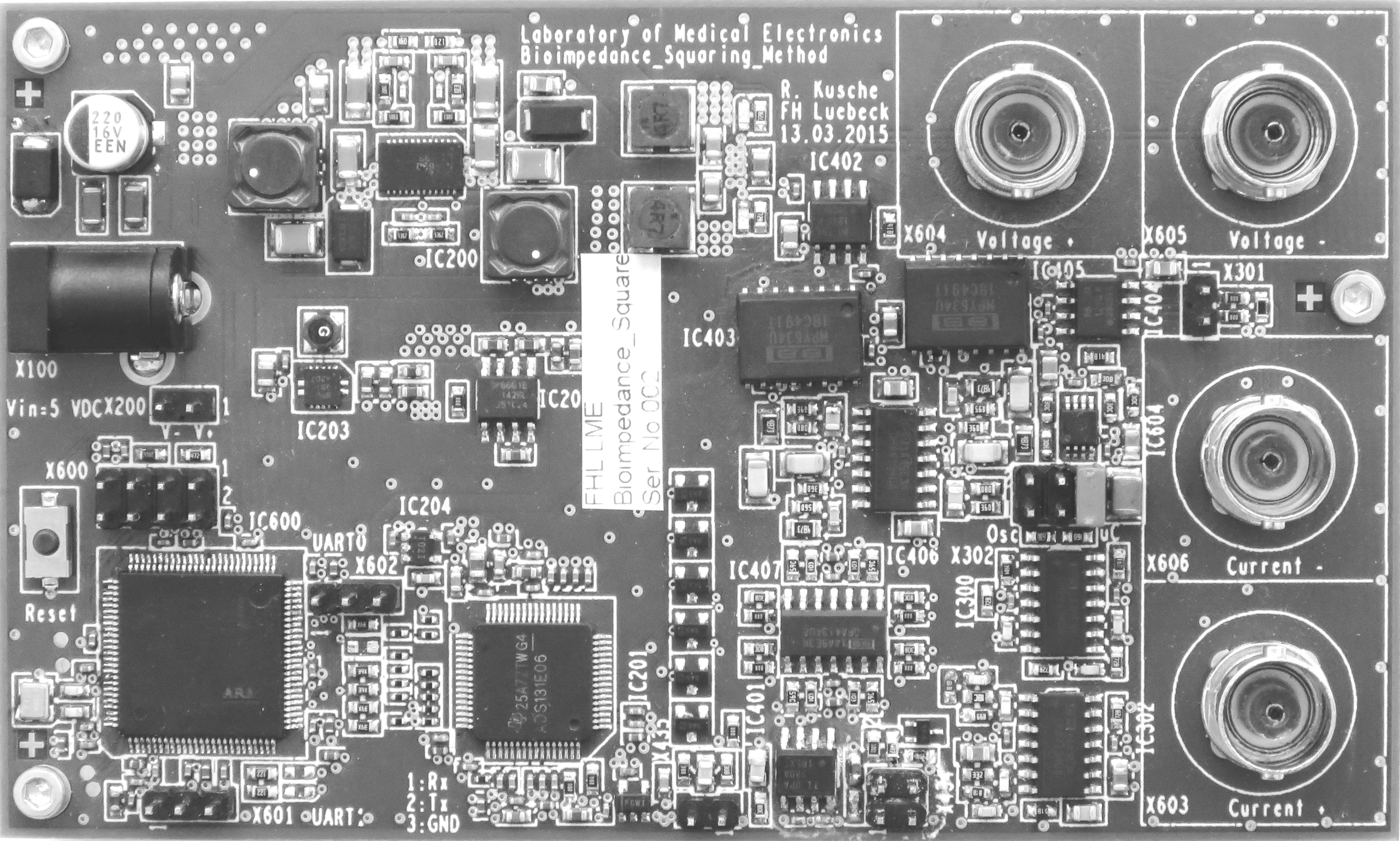}
	\caption{Photograph of the bioimpedance measurement device for pulse wave detection. The PCB has four conductive layers and dimensions of 117 mm x 71 mm.}
	\label{fig:Photograph}
\end{figure}

The block diagram of the developed measurement system is shown in Figure \ref{fig:blockDiagram}. The microcontroller (32~Bit, f$_{clk}$=120~MHz, SAM4S16C from Atmel) pictured generates a 50~kHz sinusoid and converts it into an analog voltage signal with the internal DAC (Digital to Analog Converter). In order to remove the DC offset and harmonics of the signal, a reconstruction filter (low pass, Multiple Feedback, 4th~order, Butterworth, f$_c$=100~kHz) and a high pass filter (1st order, f$_c$=1.3~kHz) are implemented back-to-back. The filtered analog voltage signal then passes through a VCCS (Voltage Controlled Current Source, Howland topology), realized by an OPAMP (Operational Amplifier, OPA211 from Texas Instruments), producing a current signal with a current limitation of I$_{max}$=1~mA, which is in compliance with the IEC60601-1. Two electrodes (Z$_E$) inject the current signal into the subject's tissue (Z$_{tissue}$) while two additional electrodes (Z$_E$) measure the voltage drop across the tissue.
\\
The differential voltages over the tissue impedance and a shunt resistor are amplified by an INA (Instrumentation Amplifier, INA128 from Texas Instruments). Afterwards, the two signals are rectified using a multiplier IC (Integrated Circuit, MPY634 from Texas Instruments) in a squaring configuration. Once squared, the signals are filtered by an active 4th~order low pass filter (Multiple Feedback, Butterworth, f$_c$=30~Hz) and then digitized by a 6~channel 24~bit ADC (Analog to Digital Converter, ADS131E06 from Texas Instruments) at 1000~samples per second. The output of a PPG (Photoplethysmography) circuit is also synchronously digitized. This allows an optical detection of the pulse wave for comparison purposes \cite{allen}. \\
The microcontroller receives the measurement data from the ADCs and sends it via a USB (Universal Serial Bus) interface to a PC to calculate the impedance magnitude and for further digital signal processing. It also provides a real time GUI (Graphical User Interface).\\
The measurement system can be powered by a medical power supply or a 3.7~V lithium-ion battery.
A photograph of the system is shown in Figure \ref{fig:Photograph}. The PCB's dimensions are 117~mm~x~71~mm and it contains about 300~components. The layout consists of four conductive layers.

\section{RESULTS}
\begin{figure}[H]
	\centering
	\includegraphics[width = 0.485\textwidth]{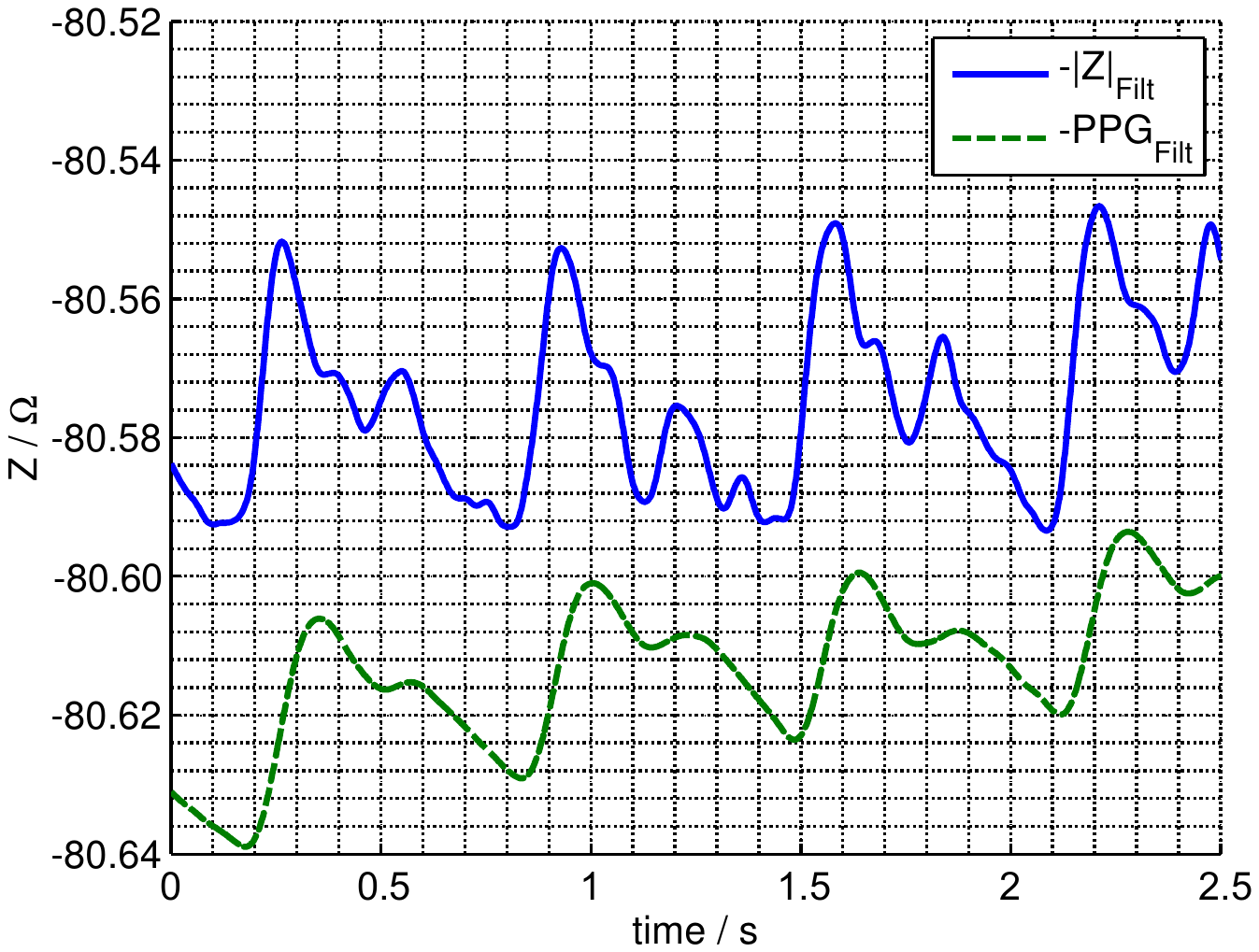}
	\caption{First measurement results executed with the developed microcontroller based system on the forearm af a young healthy male subject.}
	\label{fig:Results}
\end{figure}

To verify the realization of the method, first measurements on the forearm of a young, healthy male subject were executed.
The distance between the voltage electrodes was 10~cm. A PPG sensor was attached in between to achieve a second, independent source of information about the blood pulsation. The bioimpedance excitation signal was an AC current of 1~mA with a frequency of 50~kHz.
After acquiring the raw data, a digital filter (IIR, 2nd order, f$_C$=10~Hz), implemented in The Mathworks MATLAB, was used to remove the noise from the impedance and the PPG signal.
Figure \ref{fig:Results} displays the first measurement results over time. Both signals were inverted for better visualization.
The Z axis indicates the bioimpedance and the PPG in Ohms and arbitrary units respectively.
It can be seen, that the change in impedance is in a range of $\approx$40~m$\Omega$ with every heart beat. The morphology of the PPG signal and the impedance signal look quite similar, whereat the impedance signal seems to contain additional higher frequency components.

\section{SUMMARY AND OUTLOOK}
The results produced by this device confirm that the developed bioimpedance measurement system can be used to effectively detect the pulse wave.
First measurements proved that the arrival of the pulse wave as well as its morphology can be acquired.
In future research the device has to be tested on different subjects and the influence of motion artifacts has to be analyzed.

\section*{Acknowledgment}
The authors would like to thank Linear Technology and Texas Instruments for the provision of free samples during the development process.
\\
This publication is a result of the ongoing research within the LUMEN research group, which is funded by the German Federal Ministry of Education and Research (BMBF, FKZ 13EZ1140A/B).

\begin{figure}[h!]
\subfigure{\includegraphics[width=0.24\textwidth]{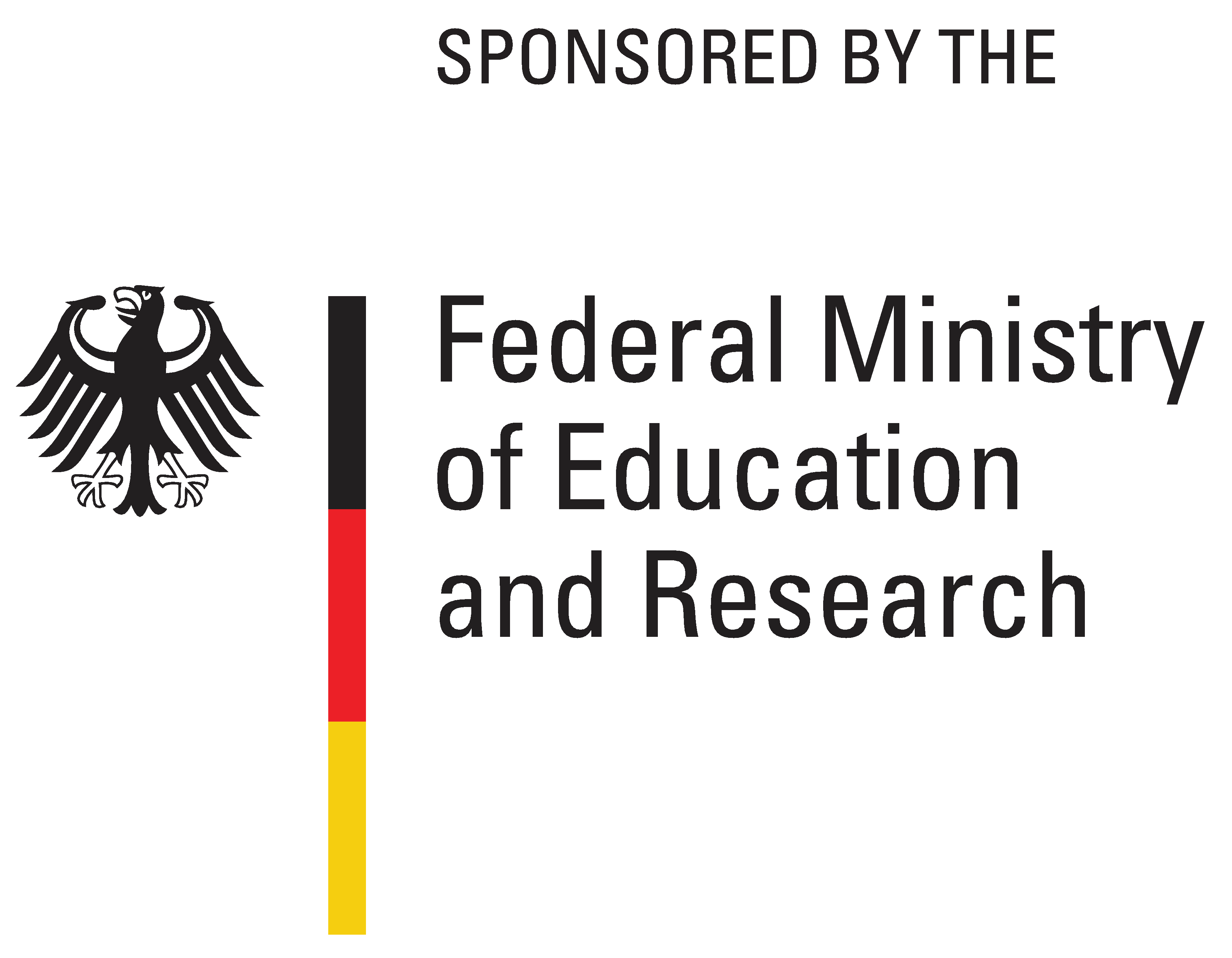}}\hfill
\subfigure{\includegraphics[width=0.24\textwidth]{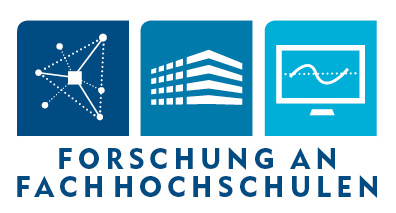}}
\end{figure}



%

\end{document}